\documentclass{aa}  

\usepackage{graphicx}
\usepackage{amsmath}
\usepackage{stmaryrd}
\usepackage{txfonts}
\usepackage{xcolor}
\usepackage{booktabs}
\usepackage{natbib}
\bibpunct{(}{)}{;}{a}{}{,} % to follow the A&A style
\usepackage[colorlinks=True, 
            urlcolor=blue,
            linkcolor=blue,
            citecolor= blue,
            dvips]{hyperref}
\usepackage{wasysym}
\newcommand{\disp}[1]{\displaystyle #1}
\newcommand{\dnz}[1]{\dfrac{d  #1}{dz}}
\newcommand{\ddnz}[1]{\dfrac{d^2  #1}{dz^2}}

\newcommand{\Amp}{{\cal A}}
\newcommand{\Fbot}{F_{\hbox{\scriptsize bot}}}
\newcommand{\Ftop}{F_{\hbox{\scriptsize top}}}

\newcommand{\Tbump}{T_{\hbox{\scriptsize bump}}}
\newcommand{\Kmin}{K_{\hbox{\scriptsize min}}}
\newcommand{\Kmax}{K_{\hbox{\scriptsize max}}}
\newcommand{\rhotop}{\rho_{\hbox{\scriptsize top}}}
\newcommand{\Ttop}{T_{\hbox{\scriptsize top}}}

\renewcommand{\na}{ \vec{\nabla} }
\newcommand{\dns}{{\hbox{\scriptsize DNS}}}

\begin{document}

\title{Direct numerical simulations of the $\kappa$-mechanism}

\subtitle{II.\ Nonlinear saturation and the Hertzsprung progression}

\author{T.\ Gastine \and B.\ Dintrans}
\institute{Laboratoire d'Astrophysique de Toulouse-Tarbes, Universit\'e
de Toulouse, CNRS, 14 avenue Edouard Belin, F-31400 Toulouse, France}

\offprints{\href{mailto:thomas.gastine@ast.obs-mip.fr}{thomas.gastine@ast.obs-mip.fr}}

\date{\today,~ $Revision: 1.99 $}

\abstract
%context
{We study the $\kappa$-mechanism that excites radial oscillations in
Cepheid variables.}
%aims
{We address the mode couplings that manage the nonlinear
saturation of the instability in direct numerical simulations (DNS).}
%methods
{We project the DNS fields onto an acoustic subspace built
from the regular and adjoint eigenvectors that are solutions to the
linear oscillation equations.}
%results
{We determine the time evolution of both the amplitude and kinetic
energy of each mode that propagates in the DNS. More than 98\% of the
total kinetic energy is contained in two modes that correspond to the
linearly-unstable fundamental mode and the linearly-stable second
overtone. Because the eigenperiod ratio is close to $1/2$, we
discover that the nonlinear saturation is due to a 2:1 resonance between
these two modes. An interesting application of this result concerns the
reproduction of Hertzsprung's progression observed in Bump Cepheids.}
%conclusion
{}

\keywords{Hydrodynamics - Instabilities - Waves - Stars: oscillations -
(Stars:variables:) Cepheids - Methods: numerical}

\maketitle

\section{Introduction}

In \cite{paper} (hereafter Paper I), we modelled the
$\kappa$-mechanism in classical Cepheids using a simplified approach,
that is, the propagation of acoustic waves in a layer of gas where the
opacity bump is represented by a hollow in radiative conductivity. To
maintain the mode excitation, the two following conditions apply:

\begin{itemize}
 \item A sufficient width and amplitude are required for the
conductivity hollow.
 \item This hollow must be located in a precise zone
called the ``transition region''.
\end{itemize}
Our parametric approach enabled us to check the reality of these two
conditions and to obtain the instability strips from a
linear-stability analysis. By starting from the
most favourable situations (i.e. the most unstable linear modes), we
performed the corresponding 1-D and 2-D nonlinear Direct
Numerical Simulations (hereafter DNS). These DNS confirmed with a
noteworthy agreement both the growth rates and spatial structures of
linearly-unstable modes. The nonlinear saturation was reached and
a quantitative study of the involved mode couplings remained to be
done.

This is the principle aim of this paper. To study the nonlinear
interactions between modes, we adapt a powerful method used in,
e.g., aeroacoustics \citep{salwen81, review-sound} or numerical
astrophysics \citep[][hereafter BCM93]{Bogdan}. It is based
on a projection of DNS fields onto a basis shaped
from the regular and adjoint eigenvectors that are solutions
to the linear oscillation equations. This method has already been used
in a simplified version by one of us to study internal
waves in DNS \citep{Dintrans04, Dintrans05}. In the quoted investigation
the eigenvalue problem was adiabatic, and therefore \textit{Hermitian},
and the authors only accounted for projections onto regular
eigenfunctions. This reduction is not applicable in our $\kappa$-mechanism
simulations since the transition region requires low densities close to
the surface and then high diffusivities (Paper I). Eigenmodes are highly
non-adiabatic and imply non-Hermitian oscillation operators with
non-orthogonal eigenfunctions. Solving the adjoint problem is therefore
mandatory to determine both mode amplitudes and energy.

By using projections onto the two respective sets of eigenvectors, the
regular and adjoint ones, the time evolution of each acoustic mode
propagating in DNS is completed. The kinetic energy in each acoustic
mode is available that highlights the energy transfer between modes. One
of the main results of this work is that more than $98\%$ of the total
kinetic energy is contained in both the fundamental mode and the second
overtone. Because this second overtone is linearly stable, we show that
its large amplitude results from a 2:1 resonance with the fundamental
mode. In our numerical experiments, this resonance is efficient because
the ratio of the two involved periods, that is $P_2 / P_0$, is close to
$1/2$.

This nonlinear saturation based on a 2:1 resonance is an interesting
result because this mechanism has been proposed to explain the secondary
bump in light curves of some classical Cepheids named ``Bump Cepheids''.
The bump position is correlated with the oscillation period that leads
to the well-known \textit{Hertzsprung progression} \citep{Hertz26,
Pay54}. The bump first appears on the descending branch of the light
curve of Population I Cepheids with periods of about 6-7 days and then
travels up this curve to reach its maximum for 10-11 day Cepheids. For
longer periods, it moves down in the ascending branch and disappears
for periods longer than 20 days \citep{Whitney83, Tsvetkov90, Bono00}.
Two main theories have been developed to explain this phenomenon:

\begin{itemize}
 \item \cite{Whitney56} suggested an ``echo mechanism'' that
\cite{Christy68} developed in more detail. In this model, two
radial-pressure waves (called Christy's waves) moving in opposite
directions are generated during compression phases in the HeII ionisation
region. The pulse running downward reflects on the stellar core
and reaches the surface at the next period during which the bump
appears.

 \item The second explanation, known as the ``resonance mechanism'', has
been proposed by \cite{Simon76} (hereafter SS76). The bump results from
a resonance between the fundamental mode and the second overtone that is
possible for a period ratio $P_2 / P_0$ close to $1/2$.

\end{itemize}
Three papers of \cite{Whitney83} and \citeauthor{Aikawa84} (1984-1985)
compared the echo and resonance theories. The basic idea was to consider
both approaches as two complementary sides of the same physical process.
Despite promising results on polytropes \citep{Whitney83}, the
dynamical approach showed that the acoustic-ray formalism did not
reproduce the ``running waves'' in Christy's diagram. This implies that
most of the energy is contained in standing waves rather than
progressive waves \citep{Aikawa84,Aikawa85}.

The resonance mechanism has also been investigated using the
amplitude-equations formalism by \cite{Goupil84} and \cite{Klapp85}.
They established the dominant role of a 2:1 resonance because phases
and amplitudes of modes show characteristic variations with the
period-ratio $P_2/P_0$ \citep[see also][]{Simon81,Kovacs89, Buchler90}.
Despite these results favoring the resonance phenomenon, the
echo-resonance controversy remains topical \citep{Fadeev92, Bono00,
Bono02}.

It is interesting to know whether our model is able to reproduce this
Hertzsprung progression. Up to 400 DNS are completed to cover a
significant range in the $P_2/P_0$ ratio, while studying the bump
position in luminosity curves. This allows us to accurately reproduce the
expected bump-progression with the ratio value: (i) if $P_2 /P_0 > 1/2$,
the bump is located in the descending branch; (ii) if $P_2 /P_0 < 1/2$,
the bump is located in the ascending branch.

In Sects.~\ref{sec:osc-model} and \ref{sec:adj}, we introduce the
general-oscillations equations and the associated adjoint problem,
respectively. We develop our projection method and provide results in
Sect.~\ref{sec:proj}. Sect.~\ref{sec:HP} concerns the Hertzsprung
progression and we outline our conclusions in
Sect.~\ref{sec:conclusion}.

\section{The pulsation model and corresponding DNS}
\label{sec:osc-model}

We focus on radial modes propagating in Cepheids and restrict our
study to the 1-D case. Our system, which represents a \textit{local}
zoom around an ionisation region, is composed of a layer of width $d$
filled with a monatomic and perfect gas with $\gamma=c_p/c_v=5/3$
($c_p$ and $c_v$ are specific heats). Gravity $\vec{g}=-g \vec{e_z}$ and
kinematic viscosity $\nu$ are assumed to be constant. Following Paper I,
the ionisation region is described by a parametric hollow in radiative
conductivity that corresponds to a bump in opacity. We recall below the
temperature-dependent profile that is adopted for the radiative
conductivity

\begin{equation}
 K_0(T)=\Kmax\left[1+\Amp\dfrac{-\pi/2+\arctan(\sigma
T^+T^-)}{\pi/2+\arctan(\sigma e^2)}\right],
\label{eq:conductivity-profile1}
\end{equation}
with

\begin{equation}
 \Amp=\dfrac{\Kmax-\Kmin}{\Kmax},\quad T^{\pm}=T-\Tbump \pm e,
\label{eq:conductivity-profile2}
\end{equation}
where $\Tbump$ is the position of the hollow in temperature and
$\sigma$, $e$ and $\Amp$ denote its slope, width and relative amplitude,
respectively.

\subsection{Instability strips from the linear-stability analysis}
\label{sec:instab}

We are interested in small perturbations about the hydrostatic and
radiative equilibria. The  layer is fully radiative and the diffusion
approximation implies that

\begin{equation}
 \vec{F}'=-K_0\na T'-K' \na T_0
\label{eq:diffus}
\end{equation}
for the radiative flux perturbation (hereafter the ``0'' subscripts mean
equilibrium quantities, while primes denote Eulerian ones).

Following Paper I, the depth $d$ of the layer is selected as the length
scale, and the top density $\rhotop$ and temperature $\Ttop$ as density
and temperature scales, respectively. The velocity scale is thus
$(c_p\Ttop)^{1/2}$. Finally, gravity and fluxes are provided in units of
$c_p \Ttop/d$ and $\rhotop (c_p \Ttop)^{3/2}$, respectively.
The linearised perturbations obey the following temperature,
momentum and continuity dimensionless equations:

\begin{equation}
 \left\lbrace
 \begin{array}{rcl}
\lambda T' &=& \dfrac{\gamma}{\rho_0}\left(K_0
\ddnz{T'}+2 \dnz{K_0}\dnz{T'}+\ddnz{K_0}T'\right) \\ \\
& & -(\gamma-1)T_0\dnz{u}+\dfrac{\Fbot}{K_0}
u,\\ \\
\lambda u &=& -\dfrac{\gamma-1}{\gamma}\left(\dnz{T'}+\dnz{\ln
\rho_0}T' +T_0 \dnz{R}\right) \\ \\
& & +\dfrac{4}{3}\nu\left(\ddnz{u}+\dnz{\ln\rho_0}
\dnz{u} \right), \\ \\\
\lambda R &=&  -\dnz{u}- \dnz{\ln \rho_0} u,
 \end{array}
 \right.
 \label{eq:syst-normal}
\end{equation}
where $R\equiv \rho'/\rho_0$ denotes the density perturbation, $u$ the
velocity and $\Fbot$ the imposed bottom flux. Here we seek normal modes
using a time-dependence of the form $\exp(\lambda t)$, with $\lambda =
\tau + \hbox{i}\omega$ ($\tau$ is the damping or growth rate of the
mode and $\omega$ its frequency). Finally, by assuming rigid walls at
both limits of the domain for the velocity, a perfect conductor at the
bottom and a perfect insulator at the top for the temperature,
we obtain the following boundary conditions

\begin{equation}
\left\lbrace\begin{array}{l}
u=0 \text{~for~} z=(0,1), \\ \\
\disp\dnz{T'}=0 \text{~for~} z=0, \\ \\
T'=0 \text{~for~} z=1.
\end{array}
\right.
\label{eq:bc-normal}
\end{equation}

Eqs. (\ref{eq:syst-normal}-\ref{eq:bc-normal}) define an eigenvalue
problem

\begin{equation}
 A \vec{\Psi}_n = \lambda_n \vec{\Psi}_n,
 \label{eq:eigenv-pb}
\end{equation}
where $\lambda_n$ is the eigenvalue associated with the (complex)
eigenvector $\vec{\Psi}_n=(T',\ u,\ R)^T$. Here $n$ defines the mode
order, that is, the number of nodes of the eigenfunction $u$ while the
matrix $A$ is a differential operator provided in
Eq.~(\ref{eq:matriceA-T'}), where the differential notation $D\equiv
d/dz$ is used for clarity.

\begin{table*}
\begin{equation}
A=
\left(
\begin{array}{cccc}
\dfrac{\gamma}{\rho_0}\left(K_0 D^2 + 2 D K_0 D + D^2K_0 \right) &
-(\gamma -1)T_0 D + \dfrac{\Fbot}{K_0} & 0 
\\ \\
-\dfrac{\gamma-1}{\gamma}(D\ln\rho_0 +D) 
& \dfrac{4}{3}\nu(D^2+D\ln\rho_0 D) & -\dfrac{\gamma-1}{\gamma} T_0 D\\ \\
0 & -D-D\ln\rho_0 & 0
\end{array}
\right) 
\label{eq:matriceA-T'}
\end{equation}
\end{table*}

\begin{figure}[t!]
 \centering
 \includegraphics[width=9cm]{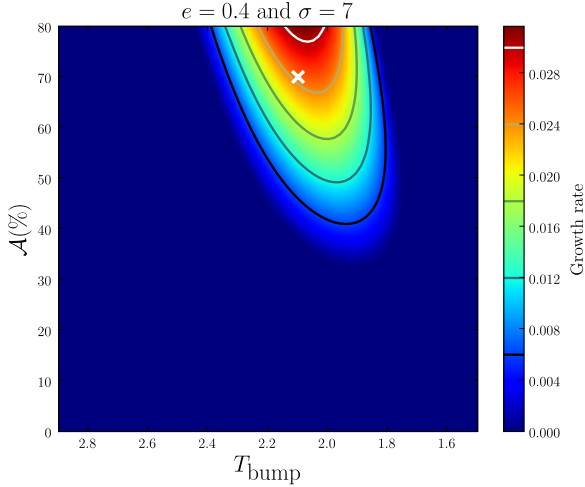}
 \caption{Instability strip for the fundamental mode in the plane
$(\Tbump,\ \Amp)$ for $e=0.4,\ \sigma=7$ and $\nu=1\times 10^{-4}$.
The top (at $z=1$) density and temperature are
$\rho=2.5\times 10 ^{-3}$ and $T=1$, respectively, while
gravity is $g=7$ everywhere. The white cross denotes a particular
simulation further studied for $\Tbump=2.1$ and $\Amp=70\%$.}
 \label{fig:IS}
\end{figure}

Figure \ref{fig:IS} displays the instability strip for the fundamental
mode\footnote{Figures are plotted using SciPy, open source
scientific tools for Python \citep{scipy}.}. In this parametric survey,
we fix the slope $\sigma$ and width $e$ of the conductivity hollow, 
whereas its amplitude $\Amp$ and temperature $\Tbump$ vary
(Eq.~\ref{eq:conductivity-profile1}). For every couple $(\Tbump,\Amp)$,
both equilibrium fields and solutions to the eigenvalue problem
(\ref{eq:eigenv-pb}) are completed using the LSB code \citep[Linear
Solver Builder,][]{LSB}. Unstable modes with $\tau > 0$ are identified
among all eigenvalues in each spectrum and only the fundamental mode is
excited by the $\kappa$-mechanism in these simulations.

\begin{table}[htbp]
\centering
\caption{First linear eigenvalues of the unstable setup used to start
the DNS (this setup is denoted by the white cross in the
instability strip in Fig.~\ref{fig:IS}). Note that only the fundamental
$n=0$ mode is
\textit{linearly} unstable.}
\begin{tabular}{cccc}
\toprule
 {\bf mode order} $\mathbf{n}$ & $\mathbf{\omega_n}$ &
$\mathbf{\tau_n\ (\times 10)}$ & period \\ \midrule \\
0 & 5.439 & $+0.250$ & 1.155 \\
1 & 8.452 & $-1.816$ & 0.743 \\
2 & 11.06 & $-4.961$ & 0.568 \\
3 & 14.36 & $-4.727$ & 0.437 \\
4 & 17.40 & $-7.825$ & 0.361 \\
5 & 20.66 & $-9.363$ & 0.304 \\
6 & 23.82 & $-11.91$ & 0.263 \\
\bottomrule
\end{tabular}
\label{table}
\end{table}

\subsection{The associated nonlinear DNS}

To confirm the reality of the instability strip and study the
mode saturation, we perform a DNS of the nonlinear problem. We start
from a linearly-unstable setup discovered in the parametric
survey (the white cross in Fig.~\ref{fig:IS} of which the oscillation
spectrum is provided in Table \ref{table}) and advance hydrodynamic 
equations in time using the Pencil-Code\footnote{See
\url{http://www.nordita.org/software/pencil-code/} and
\cite{Pencil-Code}.}.
Because we investigate the 1-D case in this work, we cannot apply the
2-D Alternate Direction Implicit (ADI) scheme developed in Paper I. We
therefore code a 1-D version of this scheme that consists of a
semi-implicit Crank-Nicholson algorithm where nonlinearities are dealt
with using a Jacobian factorisation \citep[see e.g.][\S16.6]{Numerical-Recipes}.

\begin{figure}[hbtp]
 \centering
 \includegraphics[width=9cm]{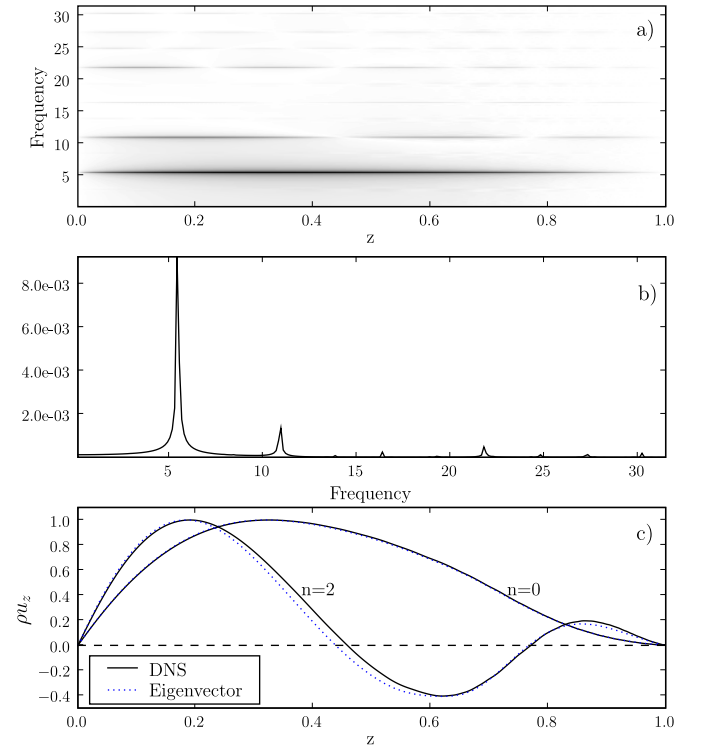}
 \caption{\textbf{a)} Temporal power spectrum for the momentum in the
$(z,\ \omega)$-plane. \textbf{b)} The resulting spectrum after
integrating over depth. \textbf{c)} Comparison between normalised momentum
profiles for $n=(0,\ 2)$ modes according to the DNS power
spectrum (solid black line) and to the linear-stability analysis (dotted
blue line).}
 \label{fig:fourier}
\end{figure}

To determine which modes are present in the DNS once it has reached the
nonlinear-saturation regime, we perform in Fig.~\ref{fig:fourier}a
a temporal Fourier transform of the momentum field $\rho u(z,t)$ and
plot the power spectrum in the $(z,\omega)$-plane. With this method,
acoustic modes are extracted because they emerge as ``shark fin
profiles'' about definite eigenfrequencies \citep[see][]{Dintrans04}. In
Fig.~\ref{fig:fourier}b, we integrate $\widehat{\rho u}(z,\omega)$ over
depth to obtain the mean spectrum. Several discrete peaks corresponding to normal
modes appear but the fundamental mode close to $\omega_0=5.439$ clearly
dominates.

The linear eigenfunctions can be compared to the mean profiles
computed from a zoom about given eigenfrequencies in the DNS power
spectrum shown in Fig.~\ref{fig:fourier}. Figure \ref{fig:fourier}c
displays such profiles about $\omega_0=5.439$ and $\omega_2=11.06$
(solid black lines), whereas associated eigenfunctions are overplotted
in dotted blue lines. The agreement between the linear-stability
analysis and the DNS is remarkable.

In summary, Fig.~\ref{fig:fourier} shows that several overtones are
present in this DNS, even for long times. Because these overtones
are linearly stable (see Table~\ref{table}), some underlying energy
transfers occur between modes. To investigate these intricate couplings
in more detail, we need to follow the evolution of each mode
\textit{separately} using the projection method developed in the next
section.

\section{The adjoint problem}
\label{sec:adj}

As said in the Introduction, the radiative diffusivity in our model is
large close to the surface due to the instability criterion (the
so-called ``transition region'' criterion). Eigenvectors
strongly differ from adiabatic ones, and therefore the quasi-adiabatic
approximation fails in this case. The dissipative effects must be fully
taken into account and the oscillations operator is non-Hermitian. In
other words, the matrix $A$ provided in Eq.~(\ref{eq:matriceA-T'}) is not
self-adjoint and its eigenvectors $\vec{\Psi_j}$ are not mutually
orthogonal. This implies that they cannot be used to arrange an acoustic
subspace on which the physical fields of the DNS are projected. We thus
consider the adjoint problem that has the same spectrum, while its
eigenvectors are orthogonal to $A$-ones because of the biorthogonality
property (see Appendix \ref{sec:app-adj}).

\subsection{The adjoint matrix $A^\dag$}

We start from the following inner-product

\begin{equation}
 \left\langle \vec{X}, \vec{Y} \right\rangle=\int_0^1 \vec{X}^\dag
\vec{Y} dz, 
  \label{eq:prod-scal}
\end{equation}
where $\dag$ denotes the Hermitian conjugate. The adjoint of the
operator $A$ is defined by \citep[e.g.][]{Lowdin83}

\begin{equation}
 \left\langle \vec{X}, A\vec{Y} \right\rangle =\left\langle A^\dag
\vec{X}, \vec{Y} \right\rangle. 
\label{eq:def_adjoint}
\end{equation}
The eigenvectors $\vec{\Phi_i}$ of $A^\dag$ are normalised to
verify (see Appen\-dix \ref{sec:app-adj})

\begin{equation}
 \left\langle\vec{\Phi_i}, \vec{\Psi_j} \right\rangle = \delta_{ij}, 
\end{equation}
where $\vec{\Phi_i}$ are the eigenvectors of $A^\dag$ such that

\begin{equation}
 \left\lbrace
 \begin{array}{rcl}
   A \vec{\Psi_j} & = & \lambda_j \vec{\Psi_j}, \\ \\
   A^\dag \vec{\Phi_i} & = & \lambda_i^* \vec{\Phi_i}.
 \end{array}
 \right.
\end{equation}

The adjoint-matrix calculation is detailed in Appendix \ref{sec:adag}
and the resulting $A^{\dag}$ is provided in
Eq.~(\ref{matrice-adjointe}). This corresponds to the following
oscillation equations for the adjoint problem:

\begin{equation}
 \left\lbrace
 \begin{array}{rcl}
  \lambda T' & = &
\dfrac{\gamma
K_0}{\rho_0}\left\lbrace\ddnz{T'}-2\dnz{\ln\rho_0}\dnz{T'}+\phantom{
\left(\dfrac{d}{dz}\right)^2}\right. \\ \\
& &\left.+ \left[-\ddnz{\ln\rho_0}+\left(\dnz{\ln\rho_0}\right)^2\right]
T'\right\rbrace \\ \\
& &+\dfrac{\gamma-1}{\gamma} \left(\dnz{u}
-\dnz{\ln\rho_0} u\right), \\ \\
\lambda u & = & (\gamma-1) T_0 \dnz{T'} +(2-\gamma)
\dfrac{\Fbot}{K_0}T'-\dnz{\ln\rho_0}R\\ \\
& &+ \dnz{R}+ \dfrac{4}{3}\nu\left[\ddnz{u}
-\dnz{\ln\rho_0}\dnz{u}
-\ddnz{\ln\rho_0} u\right ], \\ \\
\lambda R & = & \dfrac{\gamma -1}{\gamma}\left(T_0
\dnz{u}-\dfrac{\Fbot}{K_0}u\right).
 \end{array}
 \right.
 \label{eq:adj-eigenv}
\end{equation}

\subsection{The adjoint boundary conditions}

These boundary conditions are detailed in Appendix \ref{sec:adag_bc}.
While integrating Eq.~(\ref{eq:app-adjoint}) by parts, we obtain
surface contributions that must vanish to fulfill the adjoint definition
(\ref{eq:def_adjoint}) \citep{Bohlius}. This implies the following set
of adjoint boundary conditions

\begin{equation}
\left\lbrace\begin{array}{l}
u=0 \text{~for~} z=(0,\ 1), \\ \\
K_0\dnz{T'}-\left(\dnz{K_0}+K_0\dnz{\ln\rho_0}\right)T'=0
\text{~for~} z=0, \\ \\
T'=0 \text{~for~} z=1.
\end{array}
\right.
\label{eq:bc-adjoint}
\end{equation}

\subsection{Results}

\begin{figure}[h!]
 \centering
 \includegraphics[width=9cm]{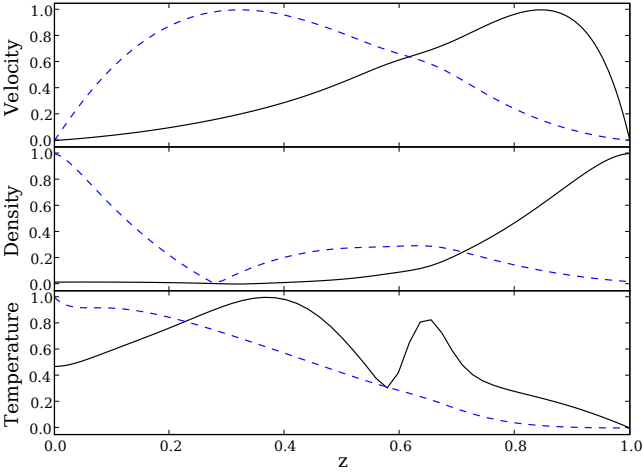}
 \caption{Real part of normalised eigenfunctions $(u,\ R,\ T')$ for
the fundamental mode (solid black line) and its adjoint (dotted blue
line), for the equilibrium setup denoted by a white cross in
Fig.~\ref{fig:IS}.}
 \label{fig:vecp}
\end{figure}

We solve the adjoint eigenvalue problem defined by
Eqs.~(\ref{eq:adj-eigenv}-\ref{eq:bc-adjoint}) again using the LSB code,
that is, we compute both the whole spectrum of eigenvalues $\lambda^*_n$
and their associated eigenvectors $\vec{\Phi_n}$. An example is given in
Fig.~\ref{fig:vecp} where we plot the real part of both the regular
(solid black line) and adjoint (dashed blue line) eigenfunctions of the
fundamental $n=0$ mode.

\section{The projection method}
\label{sec:proj}

Once the two sets of (regular) $\vec{\Psi_n}$ and (adjoint)
$\vec{\Phi_n}$ eigenvector are known, the DNS fields are projected at
each timestep using the following procedure:

\begin{enumerate}
 \item We first arrange the physical fields as 

\begin{equation}
\vec{\Xi_\dns}(z,t) = \begin{pmatrix}
           T' \\
           u\phantom{'}  \\
           R\phantom{'}
          \end{pmatrix}_\dns,
\end{equation}
where $u(z,t)$ is the velocity and $T'(z,t),\ R(z,t)$ the temperature
and density perturbations provided by

\begin{equation}
T' (z,t)= T(z,t)-T_0(z),\quad R(z,t)={\rho'\over
\rho_0}={\rho(z,t)\over \rho_0(z)}-1.
\end{equation}

\begin{table*}
\begin{equation}
 A^\dag=
 \left(
 \begin{array}{cccc}
    \dfrac{\gamma K_0}{\rho_0}\left[ D^2-2D\ln\rho_0 D
    +-D^2\ln\rho_0+(D\ln\rho_0)^2\right]
    &  \dfrac{\gamma-1}{\gamma}(D-D\ln\rho_0) & 0 \\ \\
(\gamma-1)T_0 D+(2-\gamma)\dfrac{\Fbot}{K_0}
& 
\dfrac{4}{3}\nu\left(D^2-D\ln\rho_0 D -D^2\ln\rho_0\right)
& D -D \ln \rho_0 \\ \\
0 &  \dfrac{\gamma-1}{\gamma}\left(T_0 D -\dfrac{\Fbot}{K_0}\right) & 0
 \end{array}
 \right)
 \label{matrice-adjointe}
\end{equation}
\end{table*}

 \item We decompose these fields onto the \textit{regular}-linear
eigenvectors

\begin{equation}
 \vec{\Xi_\dns}(z,t) = \Re\left\lbrace\sum_{n=0}^\infty c_n(t)
\vec{\Psi_n} (z) \right\rbrace,
\label{eq:proj}
\end{equation}
where the $c_n(t)$ coefficients are completed using \textit{adjoint}
eigenvectors  

\begin{equation}
c_n(t)=\left\langle \vec{\Phi_n},\vec{\Xi_\dns} \right\rangle = \int^1_0
\vec{\Phi_n}^\dag\ \vec{\Xi_\dns} dz.
\end{equation}

This complex function $c_n(t)$ defines the mode amplitude because
its time evolution is related to the eigenfrequency $\omega_n$ by

\begin{equation}
 c_n(t)=|c_n(t)|\ e^{\hbox{i} \phi_n(t)} \hbox{~with~}
\phi_n(t)\propto \omega_n t.
\label{eq:phik}
\end{equation}

 \item We perform Fourier transforms or phase diagrams of these
$c_n(t)$ coefficients to investigate the evolution of modes.

\end{enumerate}

\subsection{Time evolution of the complex-mode amplitudes $c_n(t)$}

\begin{figure*}[htbp]
 \centering
 \includegraphics[width=9cm]{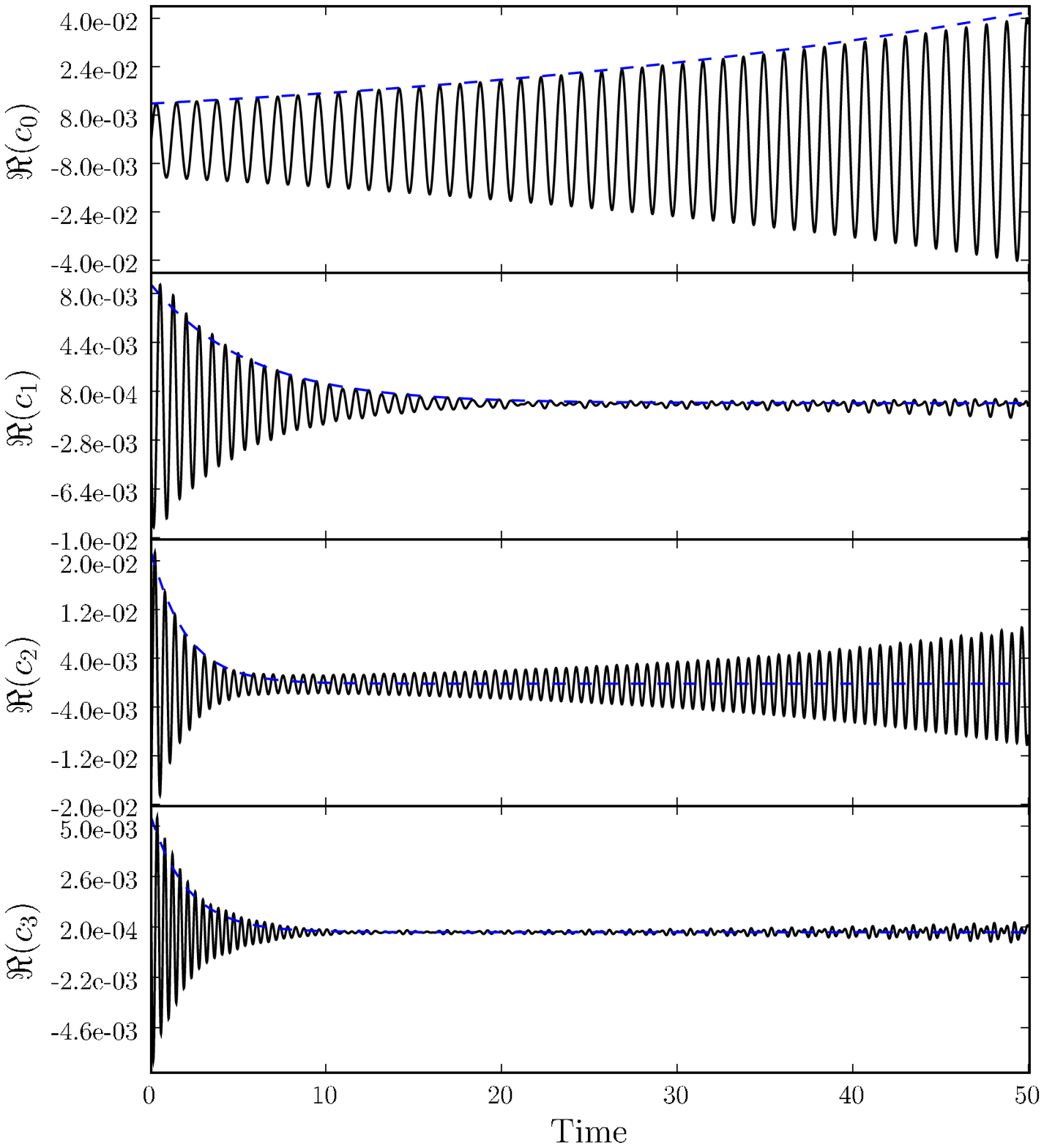}
 \includegraphics[width=9cm]{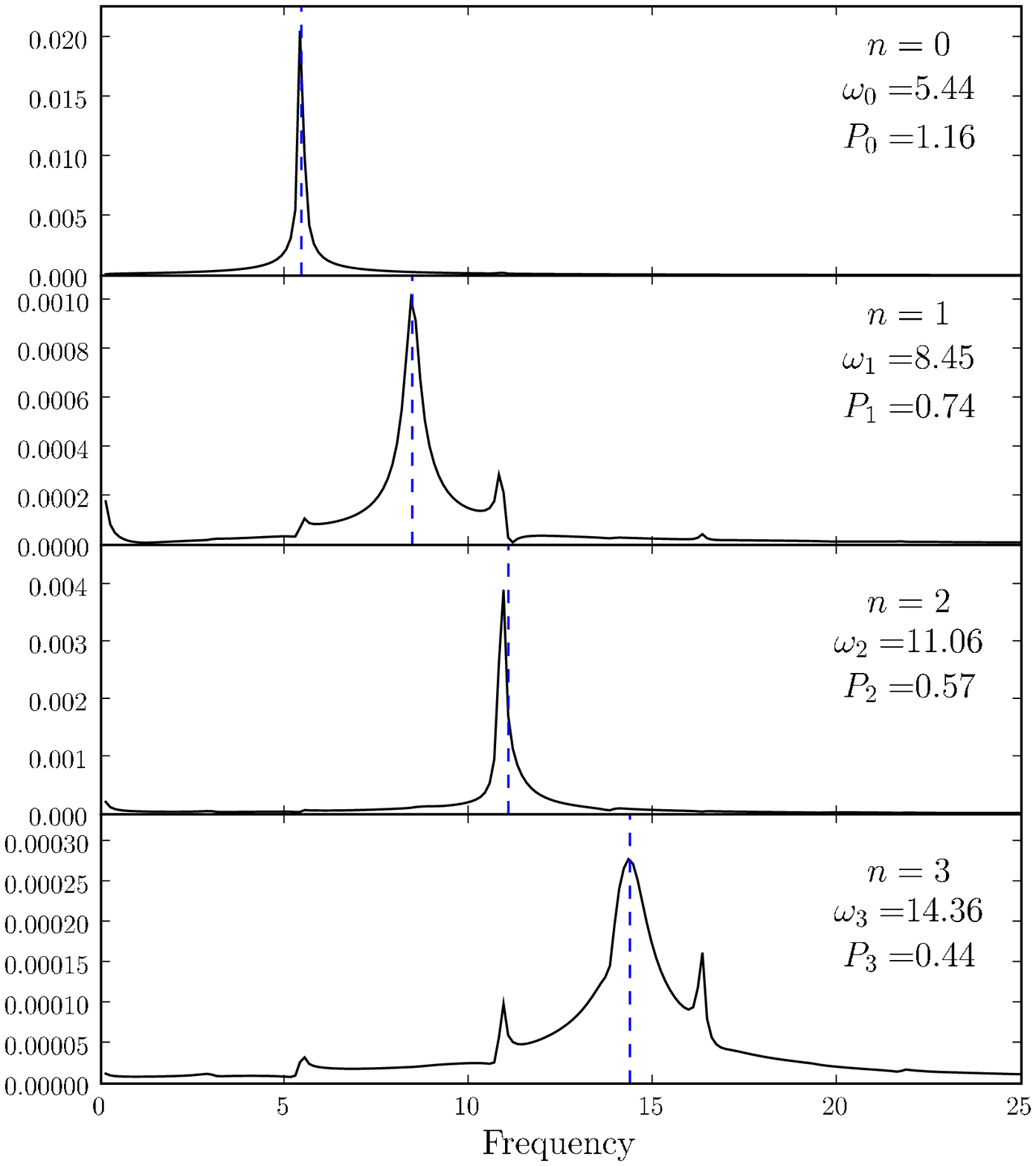}
 \caption{\textit{Left panel}: time evolution of the real part of
the complex-mode amplitudes $c_n(t)$ for $n\in \llbracket 0,\
3\rrbracket$ modes. The theoretical growth or damping curves are
overplotted in dashed blue lines. \textit{Right panel}: their
respective Fourier transforms. The modes eigenfrequencies
are overplotted in dashed blue lines.}
 \label{fig:ck}
\end{figure*}

The left panels of Fig.~\ref{fig:ck} show the time evolution of the real
part of the projection coefficients $c_n(t)$ for the first four modes
$n\in\llbracket 0,\ 3\rrbracket$. We overplot the curves $\propto
\exp(\tau_n\ t)$ derived from the linear-stability analysis (see
Table~\ref{table}). As expected, only the fundamental mode amplitude
$c_0(t)$ grows at each timestep. For other linearly-stable modes,
amplitudes decay except for the $n=2$ one. The amplitude of this
mode shows a transient decrease until time $t\simeq 10$ and then begins
to increase. This behaviour is an estimate of the nonlinear coupling that
occurs between this overtone and (at least) the fundamental mode.

We next perform Fourier transforms of these mode amplitudes (right
panel of Fig.~\ref{fig:ck}). Following Eq.~(\ref{eq:phik}), the $c_n(t)$
amplitudes behave as $c_n(t)\propto \exp(\hbox{i} \omega_n t)$ because
theoretical eigenfrequencies and peaks appearing in power spectra
overlap one another.

\subsection{Phase diagrams}

\begin{figure*}
 \centering
 \includegraphics[width=9cm]{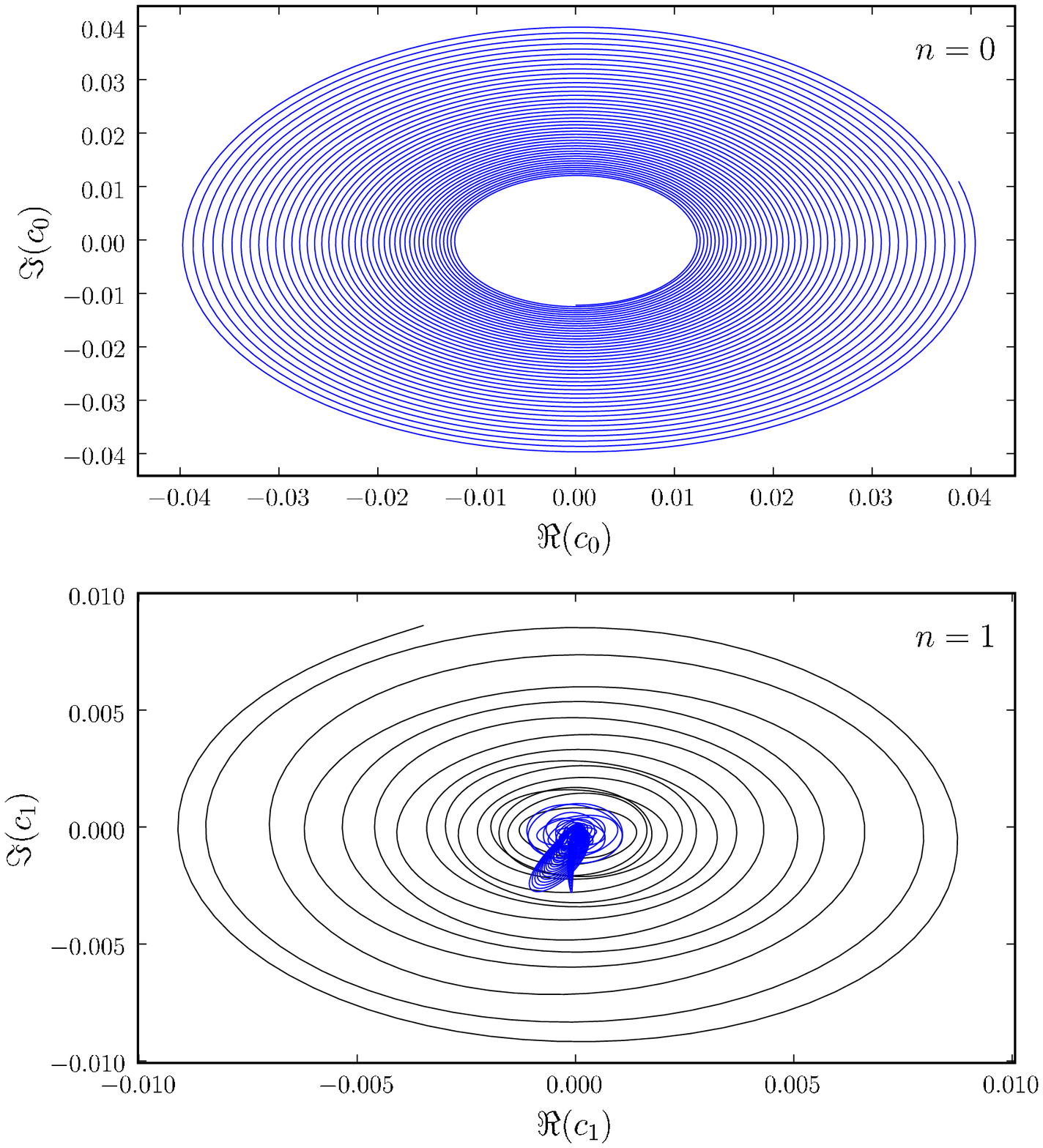}
 \includegraphics[width=9cm]{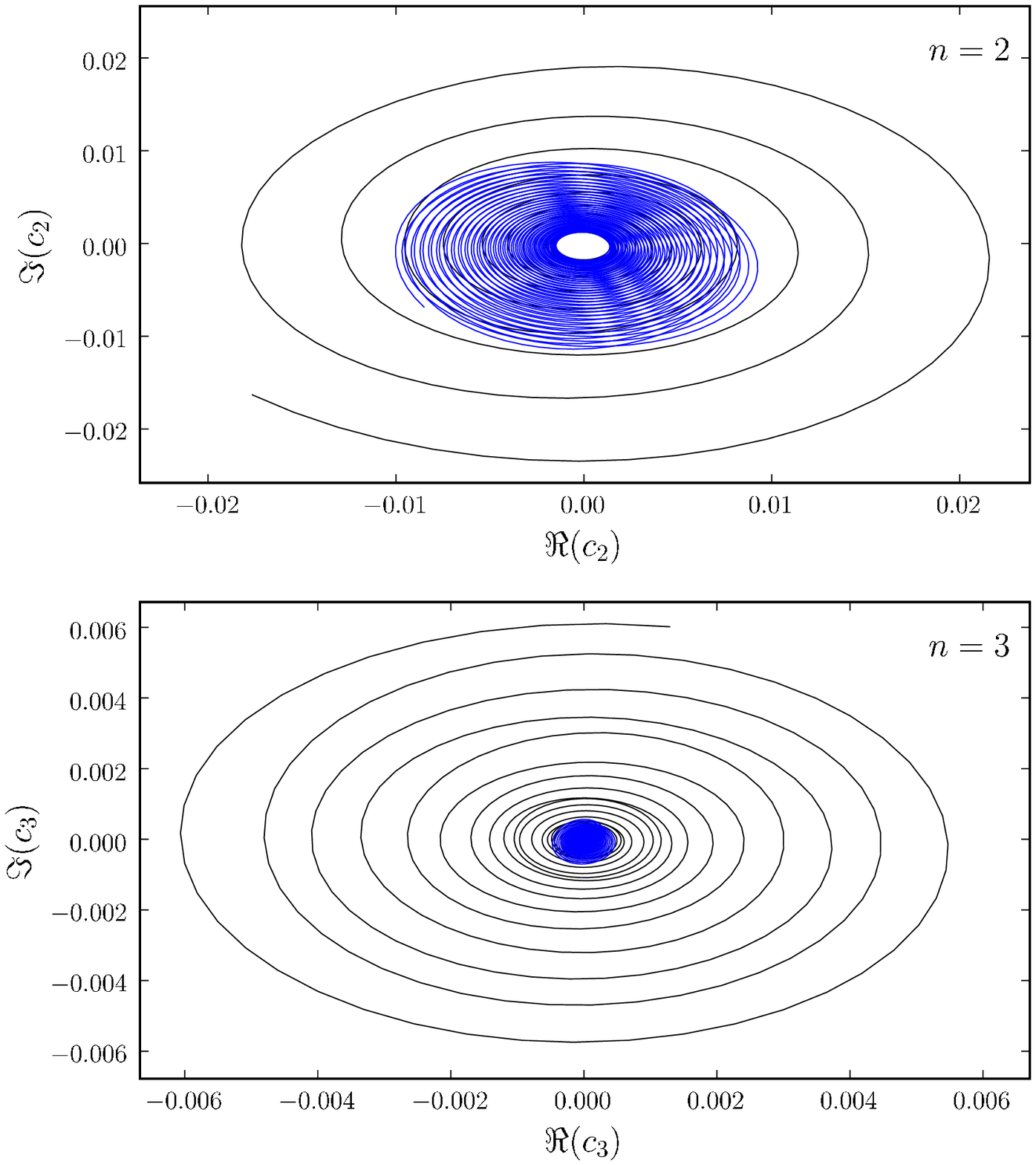}
 \caption{Phase diagrams for $n \in \llbracket 0,\ 3 \rrbracket$ modes.
The abscissa (ordinate) corresponds to the real (imaginary) part
of the complex mode amplitude $c_n(t)$. The increasing orbits are
plotted in blue, whereas the decreasing ones are in black.}
 \label{fig:phase}
\end{figure*}

In each panel of Fig.~\ref{fig:phase}, the imaginary part of the complex
mode amplitude $c_n(t)$ is plotted as a function of the real part for
the four modes shown in Fig.~\ref{fig:ck}. To avoid unreadable plots, we
stop the time evolution at $t=50$. The decreasing orbits are plotted as
black lines, whereas the increasing ones are in blue. In these diagrams,
the radius of each trajectory is related to the modulus of the complex
mode amplitude, which implies that a decreasing (increasing) radius is
associated with a damped (excited) mode. 

Once again, the fundamental mode appears to be continuously
excited. We note that the excitation phenomenon is coherent in the
sense that no discontinuity is observed in the mode orbits, that is, the
spiral develops continuously. For the second overtone, after
the linearly-transient decrease, the radius begins to increase
significantly and this is still the signature of a nonlinear coupling
with the fundamental mode. For the linearly-stable $n=1$ and
$n=3$ modes, things are more intricate because a marginal increase is
shown during the last orbits (also seen in Fig.~\ref{fig:ck}).
Unfortunately, these phase diagrams are not adapted to investigate this
long-time dynamics because orbits will overlap one another.

\subsection{The mode kinetic energy content}

To address the mode couplings responsible for the nonlinear limit-cycle
stability observed at late times, we could compute the energy content of 
each mode separately.

In BCM93, the total amount of sound generated by the turbulent
convection is weak because they find that only $0.22\%$ of
the kinetic energy is stored in acoustic standing waves. Our problem
however consists of an initially static radiative zone without
convection and the velocity field that develops is only due to
acoustic modes, that is,

\begin{equation}
 E^\text{tot}_\text{kin} = E_{\text{waves}}=\sum_{n=0}^\infty
E_n,
 \label{eq:nrj-sum}
\end{equation}
where $E_n$ is the kinetic energy contained in the $n$-acoustic mode.
One advantage of our projection method lies in its ability to compute
this kinetic energy content $E_n$ because Eq.~(\ref{eq:proj}) implies
that

\begin{equation}
 u_\dns = \Re \left\lbrace\sum_{n=0}^\infty c_n(t) u_n \right\rbrace
,
\end{equation}
where $u_\dns$ and $u_n$ are the velocity of the DNS and the regular
eigenvector, respectively. By multiplying by $1/2\rho_0 u$, the
kinetic energy density $E_{\text{kin}}(z)$ is obtained as

\begin{equation}
\begin{array}{rcl}
 E_{\text{kin}}(z) & = & \displaystyle\dfrac{1}{2}\rho_0\Re
\left\lbrace\sum_{n=0}^\infty c_n(t) u_n \right\rbrace \Re
\left\lbrace\sum_{p=0}^\infty c_p(t) u_p \right\rbrace \\ \\
& = & \displaystyle\sum_{n=0}^\infty \sum_{p=0}^\infty \dfrac{1}{2}\rho_0
\Re[c_n(t) u_n]\Re[c_p(t) u_p].
\end{array}
\end{equation}
After integrating over the domain, the total kinetic energy reads

\begin{equation}
  E^\text{tot}_{\text{kin}}=\sum_{n=0}^\infty \left(\sum_{p=0}^\infty
 \dfrac{1}{2}\int_0^1\rho_0
 \Re[c_n(t) u_n]\Re[c_p(t) u_p] dz\right).
\end{equation}
It is possible to determine the contribution of each acoustic
mode to the whole amount of kinetic energy because Eq.~(\ref{eq:nrj-sum}) allows
 us to define $E_n$ as

\begin{equation}
 E_n=\sum_{p=0}^\infty \dfrac{1}{2}\int_0^1\rho_0
 \Re[c_n(t) u_n]\Re[c_p(t) u_p] dz .
\end{equation}

\begin{figure}[htbp]
 \centering
 \includegraphics[width=9cm]{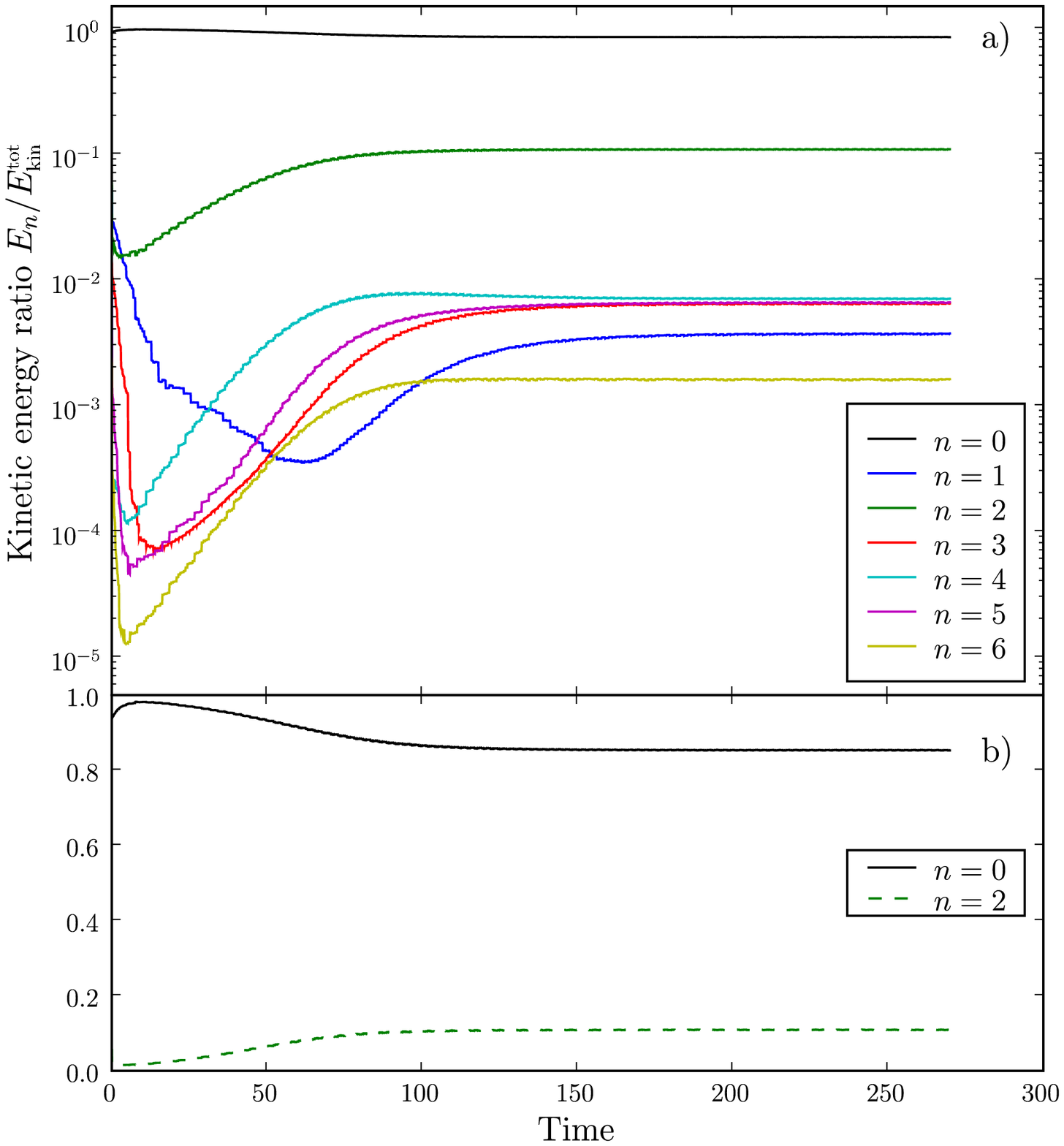}
 \caption{\textbf{a)} Kinetic energy ratios for $n\in \llbracket 0, \ 6
\rrbracket$ in a logarithmic $y$-scale. \textbf{b)} Zoom in
the $n=0$ (solid black line) and $n=2$ (dashed green line) mode energy
contents.}
 \label{fig:nrjratio}
\end{figure}

Figure \ref{fig:nrjratio}a displays the long-time evolution of the
kinetic energy ratio $E_n /E^\text{tot}_{\text{kin}}$ for $n \in
\llbracket 0, \ 6\rrbracket$. After the transient linear growth of the
fundamental mode, a given fraction of energy is progressively 
transferred to upper overtones and the nonlinear saturation is achieved 
above $t\simeq 150$. These nonlinear couplings mainly involve 
the $n=0$ and $n=2$ modes because their energy ratios are dominant.
However, the other overtones (i.e. $n\in\llbracket 1,3-6\rrbracket$)
also participate in this nonlinear saturation process but their energy
ratios remain below $1\%$. This is compatible with the weak peaks around
the corresponding frequencies $\omega_n$ shown in the power spectrum in
Fig.~\ref{fig:fourier}b or with the marginal increase of the radius
orbits in Fig.~\ref{fig:phase}. It is at first glance not surprising
because a high-order mode is associated with smaller scales and
therefore with a high damping rate (Table~\ref{table}). This implies
that the excitation of these overtones from the fundamental mode would
require efficient underlying couplings that are not observed in this
simulation.

The fact that the $n=2$ mode is more excited than the
$n=1$ one nevertheless contradicts this explanation because
its damping rate is more than twice the $n=1$ one. In
Fig.~\ref{fig:nrjratio}b, we
expand Fig.~\ref{fig:nrjratio}a for the fundamental mode and
this second overtone only. The major energy transfer well occurs between
these two modes because they contain about $98\%$ of the total kinetic
energy of this DNS. The reason for this favored coupling lies in the
period ratio existing between these two modes (see Table~\ref{table}):
the fundamental period is $P_0=2\pi/\omega_0\simeq 1.155$, while the $n=2$
one is $P_2\simeq 0.568$ such that the corresponding period ratio is
close to one half ($P_2/P_0\simeq 0.491$). The
$n=2$ mode therefore receives in a preferential way some energy from the
$n=0$ mode every two periods and that corresponds, from a dynamical point
of view, to a classical 2:1 resonance. Such a resonance is usual in
celestial mechanics with, e.g., Jupiter's moons Io ($P=1.769$d),
Europa ($P=3.551$d) and Ganymede ($P=7.154$d) and it is well
known that it helps to stabilise orbits. In our case, this
stabilisation takes the form of a nonlinear saturation: the linear
growth of the fundamental mode is balanced by the pumping of
energy from the linearly-stable second overtone behaving as an energy
sink. The final amplitudes give about $87\%$ of the kinetic energy in
the fundamental mode and $11\%$ in the second overtone, the remaining
2 percent being in higher overtones as displayed
in Fig.~\ref{fig:nrjratio}a.

\section{The Hertzsprung progression}
\label{sec:HP}

The nonlinear saturation in our excitation model therefore results from
a 2:1 resonance between the fundamental mode and the second overtone. As
shown in the Introduction, this mechanism has also been proposed to
explain the secondary bump observed in the luminosity variations of Bump
Cepheids (SS76). An interesting correlation between the value of the
ratio $P_2/P_0$ and the bump position was also emphasised leading to
the so-called ``Hertzsprung progression'' \citep{Simon81, Kovacs89,
Buchler90}. Nevertheless this correlation with $P_2/P_0$
is not observed as the period $P_2$ deduced from luminosity curves is
\textit{necessarily} equal to $P_0/2$. Indeed, the second overtone period
changes due to nonlinear couplings and finally reaches the value $P_2=
P_0/2$ once the nonlinear saturation is achieved (i.e. the 2:1 resonance). As
a consequence, only linear eigenperiods are relevant when studying the influence
of the $P_2/P_0$ ratio on the bump position. Furthermore, \cite{Buchler90} have
shown that this correlation between the phase and the $P_2/P_0$ ratio is
not reproduced in their models when plotting the phase as a function of $P_0$
only. As our hydrodynamical model deals with dimensionless quantities, studying
the phase variations according to $P_2/P_0$ is also more consistent.

\begin{figure}[htbp]
 \centering
 \includegraphics[width=9cm]{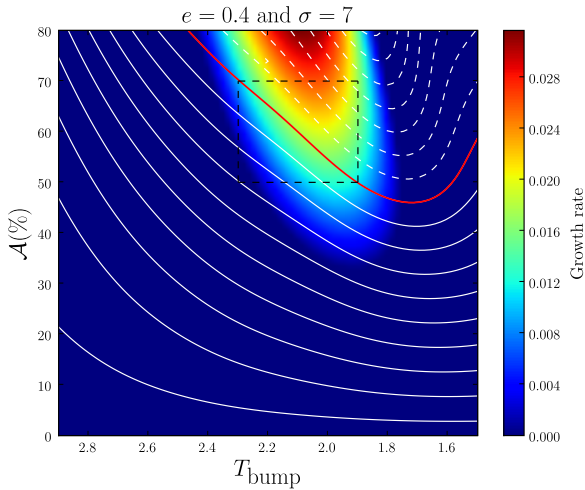}
 \caption{$P_2/P_0$ level lines overplotted on the instability strip.
Dashed white lines denote ratio $P_2/P_0 <1/2$, while solid white lines
correspond to $P_2/P_0 >1/2$. The critical level
$P_2/P_0=1/2$ is emphasised as a red line, and the dashed box delimits
the region where a large number of DNS are performed to study
Hertzsprung's progression
(see Figs.~\ref{fig:bump-position}-\ref{fig:HP}).}
 \label{fig:period}
\end{figure}

To determine whether our model is able to reproduce this
Hertz\-sprung progression, we first overplot in Fig.~\ref{fig:period}
the $P_2/P_0$ level lines with the instability strip discovered in the
linear-stability analysis (Section \ref{sec:instab}). The solid white
level lines correspond to $P_2/P_0 > 1/2$, the dashed
lines to $P_2/P_0 < 1/2$ and the red line is for the critical level
$P_2/P_0=1/2$. The instability strip is clearly split in two separate
parts and the most unstable modes belong to the $P_2/P_0<1/2$ region.
If we now perform the DNS corresponding to those different setups, we
can check if the luminosity bump position is really correlated with the
period ratio value. Because we are dealing with the 1-D case, the
luminosity reduces to the radiative flux at the top of the domain, that
is,

\begin{equation}
 L(t)\equiv\Ftop(t),
\end{equation}
and the \textit{luminosity curves} are transformed into
\textit{emerging radiative flux curves}. Because the bottom flux and the
top temperature are imposed in the DNS, these flux 
variations $\Ftop(t)$ happen about $\Fbot$.

\begin{figure}[htbp]
 \centering
 \includegraphics[width=9cm]{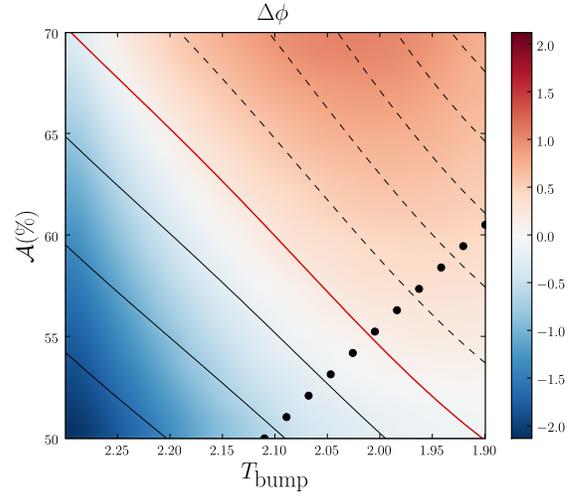}
 \caption{Isocontours of $\Delta\phi$ in the $(\Tbump,\ \Amp)$ plane
(note that we rather plot $\Delta \phi - \pi$ to highlight the critical
line $\Delta\phi=\pi$). This figure corresponds to the dashed box in
Fig.~\ref{fig:period} and the black circles display eleven
selected simulations showed in Fig.~\ref{fig:HP}.}.
 \label{fig:bump-position}
\end{figure}

We then run a large number of DNS (up to 400) in the instability strip
to cover a significant range in the $P_2/P_0$ ratio and cross the
$1/2$-line; all these DNS belong to the dashed box shown
in Fig.~\ref{fig:period}. In each case, the simulation is
performed above the nonlinear saturation ($t_{\hbox{\scriptsize
end}}=500$). Because we know that the fundamental mode and the second
overtone enter in the time evolution of the emerging flux $\Ftop(t)$, we
fit the obtained flux variations using the following decomposition:

\begin{equation}
 f(t)=\Fbot + A_0\cos(\omega_0 t + \phi_0) + A_2\cos(2\omega_0 t +
\phi_2),
 \label{eq:fit}
\end{equation}
where the free parameters $[A_0,\ A_2,\ \phi_0,\ \phi_2]$ are
computed using the Levenberg-Marquardt algorithm that is a nonlinear
least-square method \citep[e.g.][\S 15.5]{Numerical-Recipes}. The phase shift between the two signals is defined by

\begin{equation}
 \Delta \phi=\phi_2 -2 \phi_0,
 \label{eq:phase-shift}
\end{equation}
and allows first to phase the flux curves and
second, to know where the secondary bump is. If $\Delta \phi$ is
less (greater) than $\pi$, this secondary bump will be located in
the descending (ascending) branch after (before) the flux maximum.

The whole DNS are summarised in Fig.~\ref{fig:bump-position}
where isocontours of $\Delta\phi$ are plotted in the same $(\Tbump,\
\Amp)$ instability strip plane used in Fig.~\ref{fig:period}. The 
red regions are associated with $\Delta \phi  > \pi$, the blue ones
with $\Delta \phi  < \pi$ and the white ones with $\Delta \phi= \pi$.
By overplotting the level lines of the period ratio $P_2/P_0$, the
following correspondence appears:

\begin{equation}
\begin{array}{l}
\Delta \phi < \pi \hbox{~(bump \textit{after} the max.)}\Leftrightarrow P_2/P_0 > 1/2, \\ \\
\Delta \phi > \pi \hbox{~(bump \textit{before} the max.)}\Leftrightarrow P_2/P_0 < 1/2,
\end{array}
\label{eq:corresp}
\end{equation}
which corresponds to the observed correlation (SS76).

\begin{figure}[htbp]
 \centering
 \includegraphics[width=9cm]{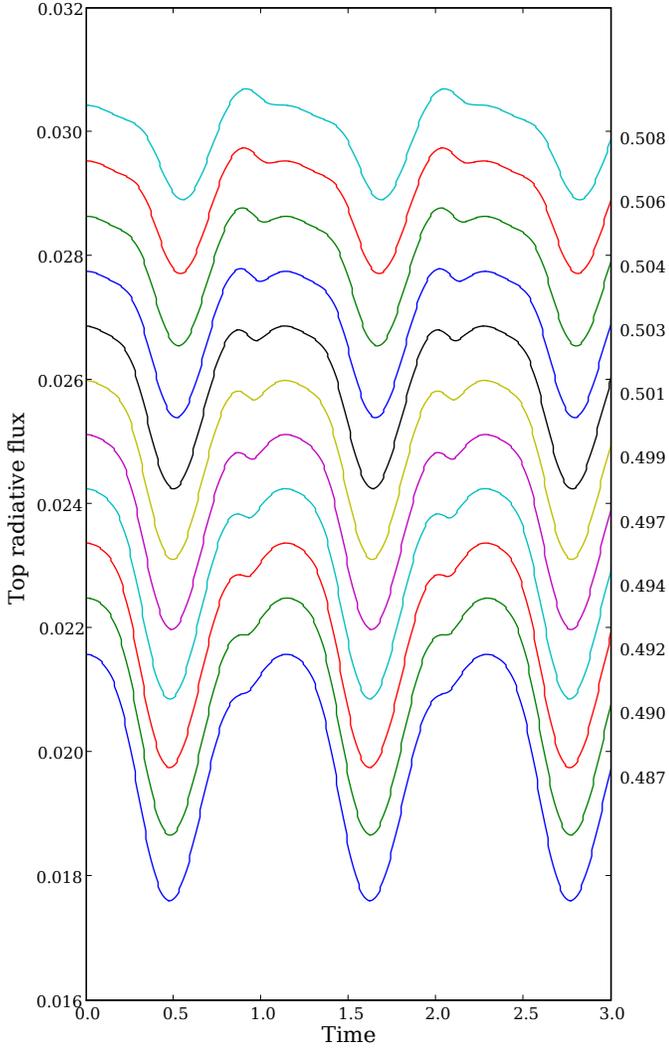}
 \caption{Rephased time evolutions of the surface radiative flux
about time $t=500$ for the eleven selected simulations plotted as black
circles in Fig.~\ref{fig:bump-position}. The ratio of eigenperiods
$P_2/P_0$ is indicated in each case.}
 \label{fig:HP}
\end{figure}

To highlight this result, examples of light curves shaping a
Hertzsprung progression are provided in Fig.~\ref{fig:HP}. These eleven
curves correspond to the black circles  distributed across the
$P_2/P_0=1/2$ level line in Fig.~\ref{fig:bump-position}.
The correspondences (\ref{eq:corresp}) appear clearly: the bump is located
in the descending branch of the light curve if $P_2/P_0 > 1/2$, while
values $P_2/P_0 <1/2$ lead to a bump in the ascending branch. 

Observations also show that the amplitude of the surface velocity
presents a \textit{double-peaked} behavior along the Hertzsprung
progression as its center corresponds to a minimum value for the velocity
\citep{Bono00}. In our simulations, no clear correlation between the
amplitude ratio $A_2/A_0$ and the bump position appears and we guess
that it may be related to the two following restrictions:

\begin{itemize}

 \item Rigid walls for the velocity have been assumed. As we are
interested in the physics involved near the top of the box, the
ratio $A_2/A_0$ is not really relevant there compared to, say, the
phase difference between the two modes.
 \item The sound speed contrast across the box is relatively weak.
As the frequency step is given by $\Delta\nu \sim \left(2 \int_L
dr/c_s\right)^{-1}$, there are not enough variations to obtain
significant changes in this ratio.
\end{itemize}

However our model is able to qualitatively reproduce the
Hertzsprung progression, as the position of the secondary bump with
regard to the main bump evolves in the expected way with the $P_2/P_0$
ratio. We therefore conclude that the 2:1 resonance supplying the
nonlinear saturation in our model results in an ``observational'' bump in
the light curve of which the position is linked to the period ratio
$P_2/P_0$.

\section{Conclusion}
\label{sec:conclusion}

In this paper, we investigate the nonlinear saturation of the acoustic
modes excited by the $\kappa$-mechanism in direct numerical simulations
(DNS). This study follows on from our former work where we began our modelling
of the $\kappa$-mechanism in Cepheids by looking the propagation of
acoustic modes in a layer of gas in which the ionisation is modelled by
a hollow in the radiative conductivity profile (Gastine \& Dintrans
2007, Paper I) 

To catch the evolution of linearly-unstable modes, we
apply an original method consisting of the projection of the DNS fields
onto the eigenmodes that are solutions to the linear oscillation
equations \citep{Dintrans04}. Because
the oscillation operator is non-Hermitian, the
regular eigenvectors are not mutually orthogonal and we consider
the adjoint problem that has the same spectrum while its
eigenvectors are orthogonal to the regular ones. Thanks to these two
sets of eigenvectors, we compute the projection coefficients for each
DNS snapshot and follow the temporal evolution of acoustic
modes separately. Corresponding phase diagrams show that
the orbit radius of the \textit{linearly}-unstable fundamental mode is
continuously growing with time. On the contrary, orbits
associated with the \textit{linearly}-stable overtones exhibit
decreasing radii, except the $n=2$ one that is increasing at
late times.

We derive the kinetic energy contributions of each mode
propagating in the DNS. More than $98\%$ of the total kinetic energy is
contained in two modes corresponding to the fundamental
mode and the second overtone. This last mode, which represents about
$10\%$ of the total kinetic energy, is
involved in the nonlinear saturation of the instability through a
2:1 resonance with the fundamental mode. It means that the unstable
fundamental mode gives energy to the stable second overtone that leads
to the limit-cycle stability, as displayed in Fig.~\ref{fig:nrjratio}.

This 2:1 resonance is striking because the same mechanism is probably
responsible for the observed bump in the luminosity curves of
Bump Cepheids, leading to the well-known ``Hertzsprung's
progression''. This progression links the bump position to the period
ratio $P_2/P_0$ existing between the fundamental mode and the second
overtone (SS76). We perform a large number of DNS covering a
significant range in this ratio and extract the resulting luminosity
curves. The 2:1 resonance does lead to an Hertzsprung progression as
the bump arises in the ascending branch for $P_2/P_0 < 1/2$, while it is
located in the descending branch for $P_2 /P_0 > 1/2$.

The study of the $\kappa$-mechanism in a purely-radiative layer is now
completed with this work. The physical criteria for the instability as
well as the nature of its nonlinear saturation are addressed. Future
works will be devoted to the influence of convection on the mode
stability. Indeed, a coupling between the convection and pulsations is
suspected to play a major role in the disappearance of unstable modes
close to the red edge of Cepheids' instability strip.
Time-dependent theories of convection have difficulty in explaining the
red-edge location since they rely on many unconstrained parameters
\citep[e.g.][]{WF98,YKB98}. Because DNS fully account for
crucial nonlinearities, they are appropriate to properly investigate 
this dynamical coupling between the acoustic modes and convection. We thus plan
to perform 2-D DNS of the $\kappa$-mechanism in the presence of convection. The
interesting cases will of course correspond to a (local)
convective-turnover timescale of the same order as the period of the
fundamental mode.

\begin{acknowledgements}
We thank the referee (G.~Bono) for his fruitful comments. Calculations were
carried out on the CalMip machine of the ``Centre Interuniversitaire de Calcul
de Toulouse'' (\url{http://www.calmip.cict.fr/}) and Grid'5000, which is an
initiative from the French Ministry of Research through the ACI GRID incentive
action, INRIA, CNRS and RENATER and other contributing partners (see
\url{https://www.grid5000.fr}).
\end{acknowledgements}

%%%%%%%%%%%%%%%%%%%%%%%%%%%%%%%%%%%%%%%%%%%
%%%%%%%%        Appendix      %%%%%%%%%%%%%
%%%%%%%%%%%%%%%%%%%%%%%%%%%%%%%%%%%%%%%%%%%

\begin{appendix}

\section{The adjoint formalism}
 
\subsection{The biorthogonality relation}
\label{sec:app-adj}

We consider a continuous linear operator $H$ with eigenvalues
$\lambda_i$ and eigenvectors $\vec{\Psi_i}$ linked by

\begin{equation}
 H\vec{\Psi_i} = \lambda_i \Vec{\Psi_i}.
\end{equation}
The operator $H^\dag$ is its adjoint with eigenvalues $\mu_i$ and
eigenvectors $\vec{\Phi_i}$ as

\begin{equation}
 H^\dag \Vec{\Phi_i} = \mu_i \Vec{\Phi_i}.
\end{equation}
Using the selected inner product (\ref{eq:prod-scal}), we have

\begin{equation}
 \langle \vec{\Psi_j}, H^\dag \vec{\Phi_i} \rangle = \mu_i \langle
\Vec{\Psi_j}, \Vec{\Phi_i} \rangle,
\end{equation}
that implies

\begin{equation}
 \langle H \vec{\Psi_j}, \vec{\Phi_i} \rangle = \mu_i \langle
\vec{\Psi_j}, \vec{\Phi_i} \rangle,
\end{equation}
by using the adjoint definition (\ref{eq:def_adjoint}). We thus obtain

\begin{equation}
 (\lambda^*_j-\mu_i)\langle \Vec{\Psi_j}, \Vec{\Phi_i} \rangle =0
 \quad \hbox{hence} \quad
\langle \Vec{\Psi_j}, \Vec{\Phi_i} \rangle = \delta_{ij}.
 \label{app-adj-valp}
\end{equation}
This property is the \textit{biorthogonality} satisfied by both the
regular and adjoint eigenfunctions \citep[e.g.][]{Lowdin83}.

\subsection{The derivation of the adjoint matrix $A^\dag$}
\label{sec:adag}

The relation (\ref{eq:def_adjoint}) that links an operator with its
adjoint takes the following form in our problem

\begin{equation}
\begin{array}{c}
 \disp\int_0^1
 \left(
 \begin{array}{ccc}
  T'_{2} & u_2 & R_2
 \end{array}\right)^*
 A\left(
 \begin{array}{c}
  T'_1  \\ u_1 \\ R_1
 \end{array}
 \right)
dz
\\ \\
=
 \disp\int_0^1
 \left(
 \begin{array}{ccc}
  T'_1  & u_1 & R_1
 \end{array}\right)
 A^\dag\left(
 \begin{array}{c}
  T'_2 \\ u_2 \\ R_2
 \end{array}
 \right)^*
dz, 
\end{array}
\label{eq:app-adjoint}
\end{equation}
where $( T'_2 \ u_2 \ R_2)^T$ is an adjoint eigenvector and $( T'_1 \
u_1 \ R_1)^T$ is a regular one. To determine the adjoint operator
$A^\dag$, we calculate each element successively from the regular
matrix in Eq.~(\ref{eq:matriceA-T'}). We provide below an example of
such a calculation for the first column of the adjoint matrix given in
Eq.~(\ref{matrice-adjointe}).

\bigskip
\noindent {\bf $\bullet$ Calculation of $A^{\dag}_{11}$}
\bigskip

\noindent $A^{\dag}_{11}$ comes from the following inner-product

\begin{equation}
\begin{array}{rcl}
I &= &\langle T'_{2}, A_{11} T'_1 \rangle  \\ \\
&= & \disp\int_0^1 T'^*_{2}\dfrac{\gamma}{\rho_0}\left[K_0 D^2+2DK_0
D +D^2
K_0\right]T'_1 dz .
\end{array}
\end{equation}
This integral is split in 3 parts as

\begin{equation}
\begin{array}{rcl}
 I &= &\gamma\left[\underbrace{\int_0^1  \dfrac{T'^*_2
K_0}{\rho_0}\ddnz{T'_1}
dz}_{(1)} +2\underbrace{\int_0^1
\dfrac{T'^*_2}{\rho_0}\dnz{K_0}\dnz{T'_1}dz}_{(2)}\right. \\ \\ 
& & \left.+ \underbrace{\int_0^1\dfrac{T'^*_2}{\rho_0}\ddnz{K_0}T'_1
dz}_{(a)}\right] .
\end{array}
\end{equation}
$(1)$ and $(2)$ are integrated by parts, leading to

\begin{equation}
 \begin{array}{rcl}
   (1) & = & \left[\dfrac{T'^*_2 K_0}{\rho_0}\dnz{T'_1}\right]_0^1
-\left[\dfrac{d}{dz}\left(\dfrac{T'^*_2 K_0}{\rho_0}\right)
T'_1\right]_0^1  \\ \\
& & +\underbrace{\displaystyle\int_0^1\dfrac{d^2}{dz^2}\left(\dfrac{T'^*_2
K_0}{\rho_0}\right) T'_1 dz}_{(b)}, \\ \\
(2) & = & 2\left[\dfrac{T'^*_2}{\rho_0}
\dnz{K_0}T'_1\right]_0^1-\underbrace{2\displaystyle\int_0^1\dfrac{d}{dz}
\left(\dfrac{T'^*_2}{\rho_0}\dnz{K_0}\right)T'_1 dz}_{(c)} .
 \end{array}
 \label{croch-lim-temp}
\end{equation}
While letting aside in a first step the integrated terms, the three
terms $(a), \ (b)$ and $(c)$ imply that
 
\begin{equation}
\begin{array}{rcl}
(a)+(b)+(c) & = &\displaystyle\int_0^1 T'_1\dfrac{\gamma
K_0}{\rho_0}\left\lbrace D^2-2D\ln\rho_0 \right. \\ \\
& & \left.-D^2\ln\rho_0+(D\ln\rho_0)^2 \right\rbrace
T'^*_{2}dz \\ \\
 & = & \langle H^\dag T'_{2}, T'_1 \rangle .
 \end{array}
 \label{eq:app-abc}
\end{equation}
Applying Eq.~(\ref{eq:app-abc}), we finally get

\begin{equation}
A^\dag_{11} = \dfrac{\gamma
K_0}{\rho_0}\left\lbrace D^2-2D\ln\rho_0 D 
-D^2\ln\rho_0+(D\ln\rho_0)^2\right\rbrace
\end{equation}

\bigskip
\noindent {\bf $\bullet$ Calculation of $A^{\dag}_{21}$}
\bigskip

\noindent $A^\dag_{21}$ comes from the following inner-product

\begin{equation}
 \langle T'_{2}, A_{12} u_1 \rangle = \int_0^1
T'^*_{2}\left[\dfrac{\Fbot}{K_0}-(\gamma-1)T_0 D\right]u_1 dz .
\end{equation}
Integrating by parts implies that

\begin{equation}
\begin{array}{rcl}
 \langle T'_{2}, A_{12} u_1 \rangle &=& \left\langle \left[(\gamma-1) T_0
D +(2-\gamma)\dfrac{\Fbot}{K_0}\right]T'_{2}, u_1\right\rangle \\ \\
&&-\underbrace{(\gamma-1)\left[ T'^*_2 T_0 u_1\right]_0^1}_{=0}.
\end{array}
\end{equation}
Using Eq.~(\ref{eq:bc-normal}), the integrated part cancels and we
obtain

\begin{equation}
 A^\dag_{21}=(\gamma-1)T_0 D +(2-\gamma)\dfrac{\Fbot}{K_0}.
\end{equation}

\bigskip
\noindent {\bf $\bullet$ Calculation of $A^{\dag}_{31}$}
\bigskip

\noindent Because $A_{13}=0$ in Eq.~(\ref{eq:matriceA-T'}), we have
 
\begin{equation}
   \langle T'_2, A_{13} R_1 \rangle = 0,
\end{equation}
and therefore

\begin{equation}
 A^\dag_{31}= 0.
\end{equation}

\subsection{The boundary conditions satisfied by $A^\dag$}
\label{sec:adag_bc}

To determine the boundary conditions for the adjoint
problem, we follow a method used in \cite{Bohlius}.
Eq.~(\ref{croch-lim-temp}) contains the boundary conditions on
temperature because the definition of the adjoint operator
(\ref{eq:def_adjoint}) imposes that every integrated term vanishes as

\begin{equation}
 \begin{array}{rcl}
 (1)+(2) &= &\left[\dfrac{T'^*_2
K_0}{\rho_0}\dnz{T'_1}\right]_0^1+2\left[\dfrac{T'^*_2 } {
\rho_0}\dnz{K_0} T'_1 \right ] _0^1 \\ \\
& & -\left[\dfrac{d}{dz} \left(\dfrac{T'^*_2 K_0}{\rho_0}\right)T'_1\right
] _0^1 =0 .
\end{array}
\end{equation}
Applying Eq.~(\ref{eq:bc-normal}), this equation implies the two
following boundary conditions on temperature:

\begin{itemize}
 \item At the top $z=1$:
 \begin{equation}
  \dfrac{T'^*_2 K_0}{\rho_0}\dnz{T'_1}=0 \Longleftrightarrow T'_2=0.
\label{eq:app-temp-bc1}
 \end{equation}
 
 \item At the bottom $z=0$:
\begin{equation}
  \left[2\dfrac{T'^*_2}{\rho_0}\dnz{K_0}-\dfrac{d}{dz}\left(
\dfrac{T'^*_2 K_0}{\rho_0} \right)\right]T'_1=0 ,
 \end{equation}
 which simplifies to
 
  \begin{equation}
  K_0\dnz{T'_2}-\left(\dnz{K_0}+K_0\dnz{\ln\rho_0}\right)T'_2=0.
  \label{eq:app-temp-bc2}
 \end{equation}
\end{itemize}

\end{appendix}

\end{document}